\documentclass[a4paper]{jpconf}
\usepackage{graphicx}

\newcommand{\be}{\begin{equation}}
\newcommand{\ee}{\end{equation}}
\newcommand{\ba}{\begin{array}}
\newcommand{\ea}{\end{array}}

\newcommand{\bea}{\begin{eqnarray*}}
\newcommand{\eea}{\end{eqnarray*}}
\newcommand{\ei}{\end{itemize}}

\begin{document}
\title{Wilson polynomials/functions and intertwining operators for the generic quantum superintegrable system on the 2-sphere}

\author{W Miller Jr and Q Li}

\address{ School of Mathematics, University of Minnesota,
 Minneapolis, Minnesota,
55455, U.S.A.}

\ead{miller@ima.umn.edu}

\begin{abstract} 
 The Wilson and Racah polynomials can be characterized as basis functions for irreducible representations of the quadratic
symmetry algebra of the quantum superintegrable system on the 2-sphere, $H\Psi=E\Psi$, with  generic 3-parameter potential. 
Clearly, the  polynomials  are expansion coefficients for one eigenbasis of a symmetry operator $L_2$ of $H$ in terms of an eigenbasis of another 
symmetry operator $L_1$, but the exact relationship appears not to have been made explicit.  We work out the details of the expansion to show, 
explicitly, how the polynomials arise and how the principal properties of these functions: the measure, 3-term recurrence relation, 2nd order difference 
equation, duality of these relations, permutation symmetry,  intertwining operators and an alternate derivation of Wilson functions --  follow from the symmetry of this quantum system. 
This paper is an exercise to show that quantum mechancal concepts and recurrence relations for Gausian hypergeometrc functions alone suffice to 
explain these properties; we make no assumptions about the structure of Wilson polynomial/functions, but derive them from quantum principles.
 There is active interest in the relation between multivariable Wilson polynomials and  the quantum superintegrable system on the $n$-sphere with 
 generic  potential, and these results should aid in the generalization.
Contracting function space realizations of irreducible representations of this quadratic  algebra  to  the other
 superintegrable systems one can  obtain the  full Askey scheme of orthogonal hypergeometric polynomials. All of these contractions of 
 superintegrable systems with potential are uniquely induced
by Wigner Lie algebra contractions of $so(3, C )$ and $e(2, C)$.  All of the polynomials produced are interpretable as quantum expansion 
coefficients. It is important to extend this 
 process to higher dimensions.
\end{abstract}

\section{Introduction}
 We define a  quantum superintegrable system  as an integrable Hamiltonian system on an $n$-dimensional pseudo-Riemannian manifold 
with potential: $ H=\Delta_n+V$   that admits $2n-1$ 
algebraically independent  partial differential operators commuting with $H$, the maximum  possible,\cite{MPW2013}.  Thus $ [H,L_j]=0,\quad n=1,2,\cdots, 2n-1$ where $\Delta_n$ is the Laplace-Beltrami operator on the manifold and we choose the generators
$L_j$ such that the sum of their orders is a small as possible.
Superintegrability captures the properties of 
quantum Hamiltonian systems that allow the Schr\"odinger eigenvalue problem $H\Psi=E\Psi$ to be solved exactly, analytically and algebraically. 
 Typically, the basis symmetries  $L_j$ generate an algebra under commutation, not usually a Lie algebra  that closes at finite order. 
 It is this algebra that is responsible for the solvability of the quantum system.

The generic superintegrable system on the 2-sphere, system $S9$ in our listing  \cite{7}, is 
 \[ H=J_1^2+J_2^2+J_3^2+\frac{a_1}{s_1^2}+\frac{a_2}{s_2^2}+\frac{a_3}{s_3^2},\qquad  a_j=\frac14-k_j^2 , \]
where      $J_3=s_1\partial_{s_2}-s_2\partial_{s_1}$  and $J_2,J_3$
are obtained by cyclic permutations of indices. Here, $  s_1^2+s_2^2+s_3^2=1$. The  operators $J_k$ preserve the order of homogeneity when acting on functions in ${\cal R}^3$ so they  act on $C^\infty$ functions on the sphere.
The basis symmetries can be chosen as  
\[L_1=J_1^2+\frac{a_3 s_2^2}{s_3^2}+\frac{a_2 s_3^2}{s_2^2},\
 L_2=J_2^2+\frac{a_1 s_3^2}{s_1^2}+\frac{a_3 s_1^2}{s_3^2},\ L_3=J_3^2+\frac{a_2 s_1^2}{s_2^2}+\frac{a_1 s_2^2}{s_1^2},\] 
 where $J_1,J_2,J_3$ are rotation generators: $J_3=s_2\partial_{s_1}-s_1\partial_{s_2}, \cdots$.
Note the discrete symmetry of $ H=L_1+L_2+L_3+a_1+a_2+a_3$ with respect to  permutations of the indices $1,2,3$. 
With the commutator $  R=[L_1,L_2]=[L_2,L_3]=[L_3,L_1]$, the  algebra generated by these symmetries has the structure
\be\label{Closure}
\begin{small}
[L_i,R]=4\{L_i,L_k\}-4\{L_i,L_j\}- (8+16a_j)L_j + (8+16a_k)L_k+ 8(a_j-a_k),
\end{small}\ee
where $i,j,k=1,2,3$, pairwise distinct. Here, $\{A,B\}=AB+BA$,  and $\{A,B,C\}$ is the symmetrizer of 3 operators. $R^2$ is contained in the algebra and expressed by
\begin{small}
\be\label{Casimir}R^2=\frac83\{L_1,L_2,L_3\} -(16a_1+12)L_1^2 -(16a_2+12)L_2^2  -(16a_3+12)L_3^2+\ee
\[\frac{52}{3}(\{L_1,L_2\}+\{L_2,L_3\}+\{L_3,L_1\})+ \frac13(16+176a_1)L_1
+\frac13(16+176a_2)L_2 + \frac13(16+176a_3)L_3 \]
\[+\frac{32}{3}(a_1+a_2+a_3)
+48(a_1a_2+a_2a_3+a_3a_1)+64a_1a_2a_3.\]
 \end{small}
  This algebra can be matched with $QR(3)$, the structure algebra of the Racah and Wilson polynomials, 
 \cite{KMP2007, 28,29,30, 31, GVZ}. The significance of these special functions is that they are the expansion coefficients of a basis of eigenfunctions 
 of one of the spherical coordinate generators 
 $L_i$ in terms of an eigenbasis of $L_j$, $i\ne j$. Long before, Dunkl in the remarkable paper \cite{Dunkl1984}, 
 (see also \cite{Koornwinder1986,Koornwinder1988} and 
 references contained therein) had computed these coefficients as expansions of polynomial bases (not eigenbases) and shown them to be 
 Racah/Wilson polynomials. However, to our knowledge no one has worked out the details of this relationship between the functions and the expansion coefficients
 to understand how important properties of Wilson polynomials can be interpreted from a quantum mechanical viewpoint.
 Here we show the critical role of intertwining operators for the quantum system in determining parameter changing recurrences for
 Wilson/Racah polynomials. In particular, the ${}_4F_3$ expressions for Wilson polynomials/functions
 follow from intertwining operators alone.

We recall the definition of ${}_4F_3$ hypergeometric functions:
 \[ {}_4F_3\left(\ba{llll} a_1,&a_2,&a_3,&a_4\\ b_1,&b_2,&b_3\ea ;x\right)=\sum_{k=0}^\infty \frac{(a_1)_k(a_2)_k(a_3)_k(a_4)_k}{(b_1)_k(b_2)_k(b_3)_k k!}\ x^k\]
where $(a)_0=1,\quad (a)_k=a(a+1)(a+2)\cdots (a+k-1)\ {\rm if}\ k\ge 1$.
 If $a_1=-n$ for $n$ a nonnegative integer then the sum is finite with $n+1$ terms. The 
Wilson polynomials of order $n$ in $t^2$ are 
\[\Phi_n(\alpha,\beta,\gamma,\delta;t)={}_4F_3\left(\ba{llll} -n,&\alpha+\beta+\gamma+\delta+n-1,&\alpha-t,&\alpha+t\\ 
\alpha+\beta,&\alpha+\gamma,&\alpha+\delta\ea ;1\right).\]
  If $\alpha+\beta=-m$, $m$ a nonnegative integer  we have the finite set $\Phi_0,\cdots,\Phi_m$ 
  of Racah polynomials.
  \section{Structure algebra and interbasis expansion coefficients}
  The basis simultaneous  eigenfunctions of $L_1$ and $H$: take the form 
\begin{small}
\[ \Psi_{N-n,n}=
(s_1^2+s_2^2)^{\frac12(2n+k_1+k_2+1)}(1-s_1^2-s_2^2)^{\frac12(k_3+\frac12)}(\frac{s_2^2}{s_1^2+s_2^2})^{\frac12(k_2+\frac12)}(\frac{s_1^2}{s_1^2+s_2^2})^{\frac12(k_1+\frac12)}\]
\[\times P_n^{(k_2,k_1)}(\frac{s_1^2-s_2^2}{s_1^2+s_2^2})P_{N-n}^{(2n+k_1+k_2+1,k_3)}(1-2s_1^2-2s_2^2),\]
\[ \frac14( L_1+2k_1k_2+2k_1+2k_2+\frac32)\Psi_{N-n,n}=-n(n+k_1+k_2+1) \Psi_{N-n,n},\quad n=0,1,\cdots, N,\]
\[  H\Psi_{N-n,n}=E_N\Psi_{N-n,n},\quad E_N = -(2N+k_1+k_2+k_3+2)^2+ \frac14,\quad N=0,1,\cdots,\]
\end{small} 
\noindent separable in spherical coordinates, $x=\cos(2\varphi),y=\cos(2\theta)$ where $s_1=\sin\theta\cos\varphi, s_2=\sin\theta \sin\varphi,s_3=\cos\varphi$, orthogonal with respect to area measure on the 1st octant of the 2-sphere.  The dimension of eigenspace $E_N$ is $N+1$.
\begin{small}
\[ P^{(\alpha,\beta)}_n(y)=\left(\ba{c} n+\alpha\\ n\ea\right){}_2F_1\left(\ba{ll} -n&\alpha+\beta+n+1\\\alpha+1\ea;\frac{1-y}{2}\right),\quad {\rm  Jacobi\ polynomials}\]
\end{small}
Define the functions $\Lambda_{N-n,n}$ by the  permutation  $1\leftrightarrow 3$, in $\Psi_{N-n,n}$. Thes are eigenfunctions of $L_2,H$:
\begin{small}
\[  \frac14(L_2+2k_2k_3+2k_2+2k_3+\frac32)\Lambda_{N-q,q}=-q(q+k_2+k_3+1) \Lambda_{N-q,q},\quad q=0,1,\cdots, N,\]
\[ H\Lambda_{N-q,q}=E_N\Lambda_{N-q,q},\quad E_N = -(2N+k_1+k_2+k_3+2)^2+ \frac14,\quad N=0,1,\cdots.\]
\end{small} \noindent
They are separable in a different set of spherical coordinates $X,Y$ expressible in terms of $x,y$ by $X=\frac{1+x+3y-xy}{xy-x+y+3},Y=\frac{x-y-1-xy}{2}$, 
and orthogonal with respect to area measure on the 1st octant of the 2-sphere. 
Due to the $S_3$ permutation symmetry there is also a basis of eigenfunctions of $L_3$ and $H$,
but we will not consider it here. 
Using the method described in \cite{KKM10c} we can derive the structure algebra of $S9$, just from the 1st order 
Gaussian differential recurrences for 
 \[ {}_2F_1\left(\ba{cc} a,&b\\c&\ea;z\right)\rightarrow  {}_2F_1\left(\ba{cc} a\pm 1,&b\\c&\ea;z\right),\quad
{}_2F_1\left(\ba{cc} a,&b\\c&\ea;z\right)\rightarrow  {}_2F_1\left(\ba{cc} a\pm 1,&b\mp 1\\c\mp 1&\ea;z\right)\]
and a limiting process.
One consequence  from Section 5.1 of that paper, is: \begin{small}
\[ L_2\Psi_{m,n}=A_n\Psi_{m-1,n+1}+B_n\Psi_{m,n}+C_n\Psi_{m+1,n-1}=\]
\[-\frac{4(N+k_3-n)(N+n+k_1+k_2+k_3+2)(n+1)(n+k_1+k_2+1)}{(2n+k_1+k_2+2)(2n+k_1+k_2+1)}\Psi_{m-1,n+1}\]
\[[-\frac{(k_1^2-k_2^2)\left(k_3^2-(2N+k_1+k_2+k_3+2)^2\right)}{2(2n+k_1+k_2+2)(2n+k_1+k_2)}
+\frac12(2n+k_1+k_2+1)^2+\frac14-k_1^2-\frac12(2N+2+k_1+k_2+k_3)^2\]
\[ +\frac12 k_3^2]\Psi_{m,n}
-\frac{4(N-n+1)(N+n+k_1+k_2+1)(n+k_1)(n+k_2)}{(2n+k_1+k_2)(2n+k_1+k_2+1)}\Psi_{m+1,n-1},\] \end{small}
 The action of $L_1$ on the $L_2$ eigenbasis follows immediately from permutation symmetry. Now we expand the $L_2$ eigenbasis in 
terms of the $L_1$ eigenbasis:
\[ \Lambda_{N-q,q}^{(k_1,k_2,k_3)}=\sum_{n=0}^N R_q^n(k_1,k_2,k_3)\,\Psi_{N-n,n}^{(k_1,k_2,k_3)},\quad q=0,\cdots. N\]
Using the self-adjoint properties of $L_1,L_2$ we can find recurrences to compute the norm:
\begin{scriptsize}\[  ||\Psi_{N-n,n}||^2=
\frac{1}{4n!\Gamma(N-n+1)}\frac{\Gamma(n+k_1+1)\Gamma(n+k_2+1)\Gamma(N-n+k_3+1)\Gamma(N+n+k_1+k_2+2)}
{(2N+k_1+k_2+k_3+2)(2n+k_1+k_2+1)\Gamma(n+k_1+k_2+1)\Gamma(N+n+k_1+k_2+k_3+2)},\]\end{scriptsize}
The squared norm of $\Lambda_{N-n,n}$ follows by applying the permutation $1\leftrightarrow 3$ to the above expression.
To better exploit the symmetry of our system we rescale the bases and expansion: \begin{scriptsize}
\[ {\Psi'}_{N-n,n}^{(k_1,k_2,k_3)}=\frac{(-1)^n n! \Gamma(N-n+1)}{\Gamma(N-n+k_3+1)\Gamma(n+k_2+1)}{\Psi}_{N-n,n}^{(k_1,k_2,k_3)},\ \
{\Lambda'}_{N-q,q}^{(k_1,k_2,k_3)}=\sum_{n=0}^N {R'}_q^n(k_1,k_2,k_3)\,{\Psi'}_{N-n,n}^{(k_1,k_2,k_3)},\quad q=0,\cdots. N\]
 \end{scriptsize}

The two sets of $N+1$ basis vectors $\{\frac{\Psi'_{m,n}}{||\Psi'_{m,n}||}\}$,  $\{\frac{\Lambda'_{p,q}}{||\Lambda'_{p,q}||}\}$ are each orthonormal, 
 implying that the 
$(N+1)\times (N+1)$ matrix
$ \left( \frac{||\Psi'_{m,n}|| {R'}^n_q}{||\Lambda'_{p,q}||}\right),\quad 0\le n,q\le N$,
is orthogonal.  We have  identities
\[\sum_{\ell=0}^N \frac{{R'}^{n_1}_\ell {R'}^{n_2}_\ell }{||\Lambda'_{N-\ell,\ell}||^2}=\frac{\delta_{n_1,n_2}}{||\Psi'_{N-n_1,n_1}||^2},\quad 
\sum_{\ell=0}^N \frac{{R'}^{\ell}_{q_1}{R'}^{\ell}_{q_2}}{||\Lambda_{N-q_1,q_1}||^2}
=\frac{\delta_{q_1,q_2}}{||\Psi'_{N-\ell,\ell}||^2}.\]
 By permutation symmetry:
$ {\Lambda'}_{p,q}^{(k_1,k_2,k_3)}={\Psi'}_{p,q}^{(k_3,k_2,k_1)}$.
We set 
\[{ R'}_q^n(k_1,k_2,k_3)_N\cdot||{\Psi'}_{N-n,n}^{(k_1,k_2,k_3)}||^2\equiv\  \Xi'\left(\ba{ccc} k_1 & k_2&k_3\\n&N&q\ea\right).\]
Note that $\Xi'$ is invariant under the transposition of its 1st and 3rd columns, and satisfies 
\[  \sum_{q=0}^N\frac{ \Xi'\left(\ba{ccc} k_1 & k_2&k_3\\n_1&N&q\ea\right) 
\Xi'\left(\ba{ccc} k_1 & k_2&k_3\\n_2&N&q\ea\right)}{||\Lambda_{N-q,q}^{(k_1,k_2,k_3)}||^2}=||{\Psi'}_{N-n_1,n_1}^{(k_1,k_2,k_3)}||^2\delta_{n_1,n_2},\]
\[ \sum_{\ell=0}^N\frac{ \Xi'\left(\ba{ccc} k_1 & k_2&k_3\\\ell&N&q_1\ea\right) 
\Xi'\left(\ba{ccc} k_1 & k_2&k_3\\\ell&N&q_2\ea\right)}{||{\Psi'}_{N-\ell,\ell}^{(k_1,k_2,k_3)}||^2}=
||{\Lambda'}_{N-q_1,q_1}^{(k_1,k_2,k_3)}||^2\delta_{q_1,q_2}.\]
Applying $L_1,L_2$ to both sides of the expansion  we have $L_1{\Lambda'}_{N-q,q}=$ $\sum_{n=0}^N\mu_n {R'}^n_q{\Psi'}_{N-n,n}$ and
\begin{scriptsize}
\[ L_2\Lambda'_{q,p}=\sum_{n=0}^N \lambda_q{R'}^n_q{\Psi'}_{N-n,n}=\sum_{n=0}^N\left({ A'}_{n-1}{R'}^{n-1}_q+
{ B'}_n{R'}^n_q+{ C'}_{n+1}{R'}^{n+1}_q\right){\Psi'}_{N-n,n},\]
\[ \quad \mu_n=-(2n+1)^2-2(2n+1)(k_1+k_2)+2k_1k_2-\frac12,\quad
 \lambda_q=-(2q+1)^2-2(2q+1)(k_2+k_3)+2k_2k_3-\frac12.\]
\end{scriptsize} By equating coefficients of $\Psi'_{N-n,n}$, we find a 3-term recurrence formula for ${R'}^n_q$, hence for $\Xi'$. 
If we make the identifications  \begin{small}
\[k_1=\delta+\beta-1,\ k_2=\alpha+\gamma-1,\ k_3=\alpha-\gamma,\ N=-\alpha-\beta,
t= q+\frac{k_2+k_3+1}{2},\] \end{small}
this formula for $\Xi'$ agrees exactly with the 3-term recurrence formula for the Racah polynomials, so, by symmetry,
$\Xi'\left(\ba{ccc} k_1 & k_2&k_3\\ n&N&q\ea\right)=K(N,k_1,k_2,k_3)R_n(k_1k_2k_3,q)$ where $R_n$ 
is proportional to a Racah polynomial in $t^2$ of order $n$ and there is permutaion 
symmetry $\left(\ba{c}k_1\\ n\ea\right)\leftrightarrow \left(\ba{c}k_3\\ q\ea\right)$.
 Moreover,
\begin{small} 
\[ ||{\Lambda'}_{N-q,q}^{(k_1,k_2,k_3)}||^2\sim \frac{\Gamma(t-\alpha+1)\Gamma(t-\beta+1)\Gamma(t-\gamma+1)
\Gamma(t-\delta+1)\Gamma(t)}{\Gamma(t+\alpha)\Gamma(t+\beta)\Gamma(t+\gamma)\Gamma(t+\delta)\Gamma(t+1)},\] \end{small}
\noindent so the measure defined by $ ||{\Lambda'}_{N-q,q}^{(k_1,k_2,k_3)}||^{-2}$ is a scalar times a function of $t$ that is
 symmetric with respect to all permutations of $\alpha,\beta,\gamma,\delta$. 
 This implies that the family of orthogonal polynomials determined by this measure must admit this symmetry up 
to a multiplicative factor. Further, the  left hand side of the orthogonality relation is proportional to \begin{small}
\[\sum_{q=0}^N \frac{(2\alpha)_q(\alpha+1)_q(\alpha+\beta)_q(\alpha+\gamma)_q(\alpha+\delta)_q}
{(\alpha)_q(\alpha-\beta+1)_q(\alpha-\gamma+1)_q(\alpha-\delta+1)_q q!}\, {\Xi'}\left(\ba{ccc} k_1 & k_2&k_3\\n_1&N&q\ea\right) 
{\Xi'}\left(\ba{ccc} k_1 & k_2&k_3\\n_2&N&q\ea\right),\]\end{small}
precisely the measure for orthogonality of the Racah polynomials $\Phi_n^{(\alpha,\beta\gamma,\delta)}(t^2)$ where $t=q+\alpha$.

{\bf Duality}: 
  By making the transpositions $k_1\leftrightarrow k_3, \ n\leftrightarrow q$ we obtain the result of applying $L_1$ to 
the expansion of $\Psi'_{N-n,n}$ in an $L_2$ eigenbasis. 
This gives a 3-term recurrence relation for $\Xi$, defining a family of orthogonal polynomials $p'_q(n)$ in the 
variable $n$. It is a 2nd order difference equation in $q$, hence $t$, for the Racah polynomials as eigenfunctions. 
This action induces a  model of an irreducible representation  of the structure algebra of $S9$ in which the basis functions are 
Racah polynomials in $t^2$ and the symmetry operators map to difference operators.

\section{Intertwining operators }
  Let ${\cal W}_{k_1,k_2,k_3}$ be the space of functions on the first octant of the 2-sphere and with 
Hamiltonian  $H^{(k_1,k_2,k_3)}$, symmetry operators $L_j^{(k_1,k_2,k_3)}$ and inner product $<\Phi,\Psi>_{k_1,k_2,k_3}$. Let ${\cal W}_{k'_1,k'_2,k'_3}$ be another such space. An {\it intertwining operator} is a mapping $X^{(k_1,k_2,k_3)}: {\cal W}_{k_1,k_2,k_3}\rightarrow
{\cal W}_{k'_1,k'_2,k'_3}$ such that
\[ X^{(k_1,k_2,k_3)}H^{(k_1,k_2,k_3)}=H^{(k'_1,k'_2,k'_3)}X^{(k_1,k_2,k_3)}.\] 
Note that $X^{(k_1,k_2,k_3)}$ maps eigenfunctions of  $H^{(k_1,k_2,k_3)}$ to eigenfunctions of 
$H^{(k'_1,k'_2,k'_3)}$ and its adjoint  $X^{*(k_1,k_2,k_3)}$  reverses the action.

Such energy shifting transformations are induced by the basic differential recurrence relations obeyed by Gaussian hypergeometric functions.
For example the standard recurrence \begin{small}
\[ \left[z(1-z) \frac{d}{dz}-(b+a-1)z+c-1\right]{}_2F_1\left(\ba{ll} a&b\\c\ea;z\right)=
(c-1){}_2F_1\left(\ba{ll} a-1&b-1\\c-1\ea;z\right),\]\end{small}\noindent
induces a 1st order differential operator \begin{small}
$ T^{(k_1,k_2,k_3)}:{\cal W}_{k_1,k_2,k_3}\rightarrow {\cal W}_{k_1-1,k_2-1,k_3}$\end{small} such that, in terms of the $x,y$ variables, \begin{footnotesize}
\[ {T^{(k_1,k_2,k_3)}}=\sqrt{1-x^2}\,\partial_x-\frac12(k_2-\frac12)\sqrt{\frac{1+x}{1-x}}
+\frac12(k_1-\frac12)\sqrt{\frac{1-x}{1+x}},\ T^{(k_1,k_2,k_3)}\Psi^{(k_1,k_2,k_3)}_{m,n}=-(n+1)\Psi^{(k_1-1,k_2-1,k_3)}_{m,n+1}.\] \end{footnotesize}

The adjoint  is induced by  
$\frac{d}{dz}{}_2F_1\left(\ba{ll} a&b\\c\ea;z\right)=
\frac{ab}{c}{}_2F_1\left(\ba{ll} a+1&b+1\\c+1\ea;z\right)$,\begin{small} \[   T^{*(k_1,k_2,k_3)}:{\cal W}_{k_1,k_2,k_3}\rightarrow 
{\cal W}_{k_1+1,k_2+1,k_3},\ \Psi^{(k_1,k_2,k_3)}_{m,n}\rightarrow
-(k_1+k_2+n+1)\Psi^{(k_1+1,k_2+1,k_3)}_{m,n-1},\]
\[{T^{*(k_1,k_2,k_3)}}=-\sqrt{1-x^2}\,\partial_x-\frac12(k_2+\frac12)\sqrt{\frac{1+x}{1-x}}
+\frac12(k_1+\frac12)\sqrt{\frac{1-x}{1+x}},\]
\end{small} Note that these intertwining operators are defined independent of basis.
The action of $T$ and $T^*$ on the $\Lambda$-basis can again be computed from 1st order relations obeyed by Gaussian hypergeometric functions.
To find these we transform to $X,Y$ coordinates and again make use of first order hypergeometric differential recurrences.
We obtain \begin{small}
\[  {T}^{(k_1,k_2,k_3)}{\Lambda'}_{p,q}^{(k_1,k_2,k_3)}=
-\frac{1}{2q+k_2+k_3+1}{\Lambda'}_{p+1,q}^{(k_1-1,k_2-1,k_3)}
 +\frac{1}{2q+k_2+k_3+1}{\Lambda'}_{p,q+1}^{(k_1-1,k_2-1,k_3)},\]
\[  { \tau}^{(\alpha-\frac12,\beta-\frac12,\gamma-\frac12,\delta-\frac12)}\
 {\Xi'}\left(\ba{ccc}k_1-1&k_2-1&k_3\\n&N+1&q\ea\right)=n(k_1+k_2+n-1)\, {\Xi'}\left(\ba{ccc}k_1&k_2&k_3\\n-1&N&q\ea\right),\] 
\[  { \tau}^{(\alpha-\frac12,\beta-\frac12,\gamma-\frac12,\delta-\frac12)}f(t)=
\frac{1}{2t}\left[ (f(t+\frac12)- f(t-\frac12)\right],\quad t=q+\frac{k_2+k_3+1}{2}.\]
 \[{ \tau^*}^{(\alpha+\frac12,\beta+\frac12,\gamma+\frac12,\delta+\frac12)}\ {\Xi'}\left(\ba{ccc}  k_1 +1& k_2+1&k_3\\ n&N-1&q\ea\right)=\,
{\Xi'}\left(\ba{ccc} k_1&k_2&k_3\\ n+1& N& q\ea\right),\]
 \[ { \tau^*}^{(\alpha+\frac12,\beta+\frac12,\gamma+\frac12,\delta+\frac12)}f(t)=
\frac{1}{2t}\left[ (\alpha+t)(\beta+t)(\gamma+t)(\delta+t) f(t+\frac12)-(\alpha-t)(\beta-t)(\gamma-t)(\delta-t) f(t-\frac12)\right].\]                                                  
                                                  \end{small}
Note that (with $t=q+\frac{k_2+k_3+1}{2}$),
\begin{small}
\[ { \tau^*}^{(\alpha+\frac12,\beta+\frac12,\gamma+\frac12,\delta+\frac12)}\ { \tau}^{(\alpha,
 \beta,\gamma,\delta)}\, {\Xi'}\left(\ba{ccc}k_1&k_2&k_3\\n&N&q\ea\right) 
=n(k_1+k_2+n+1)\,{\Xi'}\left(\ba{ccc}k_1&k_2&k_3\\n&N&q\ea\right)\] \end{small}
a 2nd order difference equation for $\Xi'$ as a polynomial in $t^2$. We will solve this equation.

 \subsection{Calculation of Racah polynomials}
  Note that $\Xi'_n(t)$ can be written as $\Xi'=G(n,N,k_1,k_2,k_3)\,\Phi_n(t)$ where $\Phi_n(t)$ is a 
  polynomial in $t^2$ such that $\Phi_n(0)=1$. Thus we can write 
      $ \Phi_n(t)=\sum_{k=0}^{n}w_k P_k(\alpha,t)$, where $ P_k(\alpha,t)=(\alpha+t)_k(\alpha-t)_k$, $w_0=1$.
      Applying  $\tau$ to  $P_k(\alpha,t)$, we get,
      \be\label{tauat}\tau P_k(\alpha,t)=-kP_{k-1}(\alpha+\frac12,t)\ee
      Applying $\tau^*$ to the shifted basis,  we get \begin{small}
\be \label{tau*at}\tau^*P_{k}(\alpha+\frac12,t)= -(\alpha+\beta+\gamma+\delta+k) P_{k+1}(\alpha,t)+(\alpha+\beta+k)(\alpha+\gamma+k)(\alpha+\delta+k)P_k (\alpha,t)_k.
\ee \end{small}
Thus,
      \[\tau^*\tau P_k(\alpha,t)= k(\alpha+\beta+\gamma+\delta+k-1)P_k(\alpha,t)
-k(\alpha+\beta+k-1)(\alpha+\gamma+k-1)(\alpha+\delta+k-1)P_{k-1}(\alpha,t), \]
 a 2-term recurrence relation, which implies
         $ w_{k+1}=\frac{(-n+k)(n+\alpha+\beta+\gamma+\delta+k-1)}{(k+1)(\alpha+\beta+k)(\alpha+\gamma+k)(\alpha+\delta+k)}w_k$, $w_0=1$. 
      It is easy to solve this recurrence to obtain 
      \be\label{omegak} w_k=\frac{(-n)_k(n+\alpha+\beta+\gamma+\delta-1)_k}{k!(\alpha+\beta)_k(\alpha+\gamma)_k
      (\alpha+\delta)_k},\ k=0,1,\cdots.\ee
    Hence, unique up to a scalar multiple,
\be\label{Phin} \Phi_n(t)={}_4F_3\left(\ba{llll} -n&n+\alpha+\beta+\gamma+\delta-1&\alpha+t&\alpha-t\\ 
\alpha+\beta&\alpha+\gamma& \alpha+\delta\ea; 1\right).\ee

\subsection{More permutation symmetry:}
 Since the measure for the Racah polynomials is invariant under all permutations of $\alpha,\beta,\gamma,\delta$, and the Racah 
polynomials can be obtained from the measure by the Gram-Schmidt process. Each such permutation must take a Racah polynomial to a scalar multiple of itself.
We find 
$ (\alpha+\beta)_n(\alpha+\gamma)_n(\alpha+\delta)_n \, 
{\Phi}^{(\alpha,\beta,\gamma,\delta)}_n(t)$
is invariant under all permutations. 

\subsection{More intertwining operators}
The recurrence  \begin{small}
 $$ \left( z\frac{d}{dz}+c-1\right) {}_2F_1\left(\ba{ll} a&b\\\ \quad c\ea;z\right)=(c-1){}_2F_1\left(\ba{ll} a&b\\\ \quad c-1\ea;z\right) $$
leads to an intertwining operator 
$ U^{(k_1,k_2,k_3)}_{(-,+,-,+)} :\ {\cal W}_{k_1,k_2,k_3}\longrightarrow {\cal W}_{k_1+1,k_2-1,k_3}$, 
\[ \mu ^{(\beta,\delta)}\, \Phi^{(\alpha-\frac12,\beta+\frac12,\gamma-\frac12,\delta+\frac12)}_n(t)=
 \frac{(n+\beta+\delta)(n+\alpha+\gamma-1)}{(\alpha+\gamma-1)}\,  \Phi^{(\alpha,\beta,\gamma,\delta)}_n(t)\] 
Its action on the $\Lambda'$ basis induces the recurrence 
\[ { \mu}^{(\beta,\delta)}f(t)=
\frac{1}{2t}\left[ (\beta+t)(\delta+t)f(t+\frac12)-(\beta-t)(\delta-t) f(t-\frac12)\right],\]
 \[ \mu ^{(\beta,\delta)}\, {\Xi'}\left(\ba{ccc} k_1+1&k_2-1&k_3\\ n&N&q\ea\right)=
 (n+k_1+1)(n+k_2)\,\, {\Xi'}\left(\ba{ccc} k_1&k_2&k_3\\ n&N&q\ea\right).\]
\end{small}
The permutation invariance  of the Racah polynomials leads to a family of recurrences in
the $\mu$ such that any pair of $\alpha,\beta,\gamma, \delta$ can be raised by $\frac12$ and 
the other pair lowered
by $\frac12$. These   also follow from intertwining operators induced by Gaussian hypergeometric differential recurrences. 
In particular, the hypergeometric recurrence \begin{small}
$$ \left( z\frac{d}{dz}+a\right) {}_2F_1\left(\ba{ll} a&b\\\ \quad c\ea;z\right)=a \ {}_2F_1\left(\ba{ll} a+1&b\\\ \quad  c\ea;z\right) $$
induces the operator 
 {\footnotesize \be \label{eqn3} U^{(k_1,k_2,k_3)}_{(+,+,-,-)}=\sqrt{ \frac{1+y}{2}}\left[ 
 -(1-y)\partial_y-N-\frac{k_1}{2}-\frac{k_2}{2}-\frac12+\frac12(k_3+\frac12)\left( \frac{1-y}{1+y}\right)     \right], \ee}
 \be  U^{(k_1,k_2,k_3)}_{(+,+,-,-)}\Psi^{(k_1,k_2,k_3)}_{m,n}=-(n+N+k_1+k_2+1)\Psi^{(k_1,k_2,k_3+1)}_{m-1,n},\quad m\ge 1,n\ge 0. \ee
The action on the $L_2$ eigenbasis is
\[ \mu ^{(\alpha,\beta)}\, {\Xi'}\left(\ba{ccc} k_1&k_2&k_3+1\\ n&N-1&q\ea\right)=-\frac{2N+k_1+k_2+k_3+2}{2N+k_1+k_2+k_3+1}
 \,\, {\Xi'}\left(\ba{ccc} k_1&k_2&k_3\\ n&N&q\ea\right),\]
\be\label{murec} \mu^{(\alpha,\beta)}\Phi_n^{(\alpha+\frac12,\beta+\frac12,\gamma-\frac12,\delta-\frac12)}=(\alpha+\beta)
\Phi_n^{(\alpha,\beta,\gamma,\delta)},\quad \alpha+\beta=-N.\ee
\end{small}

 The recurrence 
$ \left( z\frac{d}{dz}+b\right) {}_2F_1\left(\ba{ll} a&b\\\ \quad c\ea;z\right)=b \ {}_2F_1\left(\ba{ll} a&b+1\\\ \quad  c\ea;z\right)$, 
induces the operator 
 {\footnotesize \be \label{eqn5} U^{(k_1,k_2,k_3)}_{(+,-,-,+)}=\sqrt{ \frac{1+y}{2}}\left[  (y-1)\partial_y+N+\frac{k_1}{2}+\frac{k_2}{2}+k_3+\frac32+\frac12(k_3+\frac12)\left( \frac{y-1}{y+1}\right)     \right],\ee}
 \be U^{(k_1,k_2,k_3)}_{(+,-,-,+)}\Psi^{(k_1,k_2,k_3)}_{m,n}=(n+N+k_1+k_2+k_3+2)\Psi^{(k_1,k_2,k_3+1)}_{m,n},\quad m\ge 0,n\ge 0. \ee
In terms of $\Xi'$ and $\Phi_n$ the action is
\[ \mu ^{(\alpha,\delta)}\, {\Xi'}\left(\ba{ccc} k_1&k_2&k_3+1\\ n&N&q\ea\right)=
 \frac{2N+k_1+k_2+k_3+2}{2N+k_1+k_2+k_3+3}\,\, {\Xi'}\left(\ba{ccc} k_1&k_2&k_3\\ n&N&q\ea\right),\]
\[ \mu^{(\alpha,\delta)}\Phi_n^{(\alpha+\frac12,\beta-\frac12,\gamma-\frac12,\delta+\frac12)}=(\alpha+\delta)
\Phi_n^{(\alpha,\beta,\gamma,\delta)},\ \quad \alpha+\delta=N+k_1+k_2+k_3+2.\]

Note that the operators (\ref{eqn3},\ref{eqn5}) are not basis independent, since they depend on $N$. However, the 
 intertwining operator $V^{(k_1,k_2,k_3)} :\ {\cal W}_{k_1,k_2,k_3}\longrightarrow {\cal W}_{k_1,k_2,k_3+1}$,
\be\label{Vop}\ 
 V^{(k_1,k_2,k_3)}=U^{(k_1,k_2,k_3)}_{(+,-,-,+)}+ U^{(k_1,k_2,k_3)}_{(+,+,-,-)}= \sqrt{ \frac{1+y}{2}}\left[ 2 (y-1)\partial_y+k_3+1\right].\ee
is  basis independent, so defined  intrinsically, as is its adjoint.

\subsection{The expansion coefficients}
Solving all of these recurrences for $\Xi'$, we find  
\begin{small}
\[{ R'}_q^n(k_1,k_2,k_3)_N\cdot||\Psi'_{N-n,n}(k_1,k_2,k_3)||^2=
  \Xi'\left(\ba{ccc} k_1 & k_2&k_3\\n&N&q\ea\right)=\]
 \[\frac{4c\  (2N+k_1+k_2+k_3+2)\Gamma(N+1)}{\Gamma(N+k_1+k_2+k_3+2)
\Gamma(k_2+1)}\Phi_n^{(\alpha,\beta,\gamma,\delta)}(t^2),\quad t= q+\frac{k_2+k_3+1}{2},\]
\[
 \alpha=\frac{k_2+k_3+1}{2},\ 
 \beta=-N-\frac{k_2+k_3+1}{2},\ \gamma =\frac{k_2-k_3+1}{2},\ \delta=N+k_1+\frac{k_2+k_3+3}{2},\]
\[ \Phi_n=
{}_4F_{3}\left(\begin{array} {llll}-n,&k_1+k_2+n+1,&-q,&k_3+k_2+q+1\\
 -N,&k_2+1,&N+k_1+k_2+k_3+2\end{array};1\right)\]
\end{small} 
 The overall scaling factor $c(k_1,k_2,k_3)$ can be determined by evaluating the double integral for $\Xi'$ in
the simplest case $n=q=N=0$ where it factors into a product of beta integrals.

\section{Extension  to Wilson polynomials} 
 Racah polynomials are $S9$ expansion coefficients for finite dimensional representations on the real 2-sphere.  Wilson polynomials 
are expansion coefficients related to infinite dimensional representations for  the Schr\"odinger eigenvalue 
equation of the generic potential  
on the upper sheet of the 2d hyperboloid. (Much earlier, Koornwinder \cite{Koornwinder1986,Koornwinder1988} pointed out the connection 
between expansion coefficients on higher dimensional
hyperboloids but not specifically to the 2d case.)  A Hilbert space structure is imposed on the eigenspace corresponding to 
a single continuous spectrum eigenvalue,
where $N$ is a negative real number, not an integer.

We expand the $L_2$ basis vectors in terms of the $L_1$ basis:
\[\Lambda_{q}=\sum_{n=0}^\infty R_q^n\Psi_{n},\quad \sum_{\ell=0}^\infty \frac{||\Psi_{n_1}||^2 }{||\Lambda_{\ell}||^2}R^{n_1}_\ell R^{n_2}_\ell=\delta_{n_1,n_2},\]
Applying $L_2$ to both sides of the expansion we can show that the $R^n_q$ satisfy a three term recurrence relation and a difference equation as before,
and the orthogonality relation can  be rewritten in the form \begin{small}
\[ \sum_{q=0}^\infty \frac{(2\alpha)_q(\alpha+1)_q(\alpha+\beta)_q(\alpha+\gamma)_q(\alpha+\delta)_q}
{(\alpha)_q(\alpha-\beta+1)_q(\alpha-\gamma+1)_q(\alpha-\delta+1)_q\, q!}\,{R'\,}^{n_1}_q {R'\,}^{n_2}_q=\delta_{n_1,n_2}h_{n_1},\] \end{small}
 where ${R'}^n_q\sim \Phi_n(t)$ is a Wilson polynomial. This is equivalent to a ${}_5F_4$ identity and can all be made rigorous. Wilson
recast the orthogonality into the form of a contour integral which greatly extended its domain of validity. All of the Racah-intertwining operators  
extend to this case.

 The quantum problem on the 2d hyperboloid has mixed spectrum, both bound states and continuous spectra, \cite{FW}. An interesting task for future research 
 is to work out the interbasis expansion associated with the spectral decomposition in this case and to relate it explicitly to partially discrete, partially continuous 
 orthogonality relations for Wilson polynomials.

\section{Extension to Wilson functions}
If $n$ is not an integer a formal calculation using (\ref{tauat}) and (\ref{tau*at}) that ignores series convergence still gives
(\ref{Phin}) as a solution of 
the eigenvalue equation $\tau^*\, \tau\,\Phi_n=n(n+\alpha+\beta+\gamma+\delta-1)\Phi_n$.
However, a careful calculation gives
\[[\tau^*\, \tau\,-n(n+\alpha+\beta+\gamma+\delta)] \sum_{k=0}^K\omega_k\, P(t,\alpha)_k=
\frac{(-n)_{K+1}(n+\alpha+\beta+\gamma+\delta-1)_{K+1}}{K!\ 
(\alpha+\beta)_K(\alpha+\gamma)_K(\alpha+\delta)_K}(\alpha+t)_K(\alpha-t)_K. \]
Taking the limit as $K\to +\infty$, and making use of the Stirling formula, we obtain
\be\label{generalbasis1}  
\left[\tau^*\, \tau \, - n(n+\alpha+\beta+\gamma+\delta)\right]\Phi_n(t)=\frac{\Gamma(\alpha+\beta)\Gamma(\alpha+\gamma)\Gamma(\alpha+\delta)}
{\Gamma(-n)\Gamma(n+\alpha+\beta+\gamma+\delta-1)\Gamma(\alpha+t)\Gamma(\alpha-t)}.\ee
Since the $\Gamma$ function has a pole at the negative integers, we see that $\Phi_n(t)$ satisfies the eigenvalue equation for $n$ a 
nonnegative integer, but otherwise it does not, except for isolated choices of the parameters. 
 Now consider the functions with relations \begin{small}
\[ Q(t,\alpha,\beta)_k=\frac{\Gamma(1-\beta+t)\Gamma(1-\beta-t)}{\Gamma(\alpha+t)\Gamma(\alpha-t)}(1-\beta+t)_k(1-\beta-t)_k,\
\tau\  Q(t,\alpha,\beta)_k=(\alpha+\beta-k-1)\ Q(t,\alpha+\frac12,\beta+\frac12)_{k},\]
\[ \tau^*\ Q(t,\alpha+\frac12,\beta+\frac12)_k=-(\gamma+\delta +k)\ Q(t,\alpha,\beta)_k+k(k-\beta+\delta)(k-\beta+\gamma)\
 Q(t,\alpha,\beta)_{k-1}.\] \end{small}
 Then computing formally without regard to series convergence,  we obtain the nonpolynomial solution  of the $\tau^*\tau$ eigenvalue 
 equation:
 \be\label{basis2}\Psi_n(t)=\frac{\Gamma(1-\beta+t)\Gamma(1-\beta-t)}{\Gamma(\alpha+t)\Gamma(\alpha-t)}
 {}_4F_3\left(\ba{llll}1 -n-\alpha-\beta&n+\gamma+\delta&1-\beta+t&1-\beta-t\\ 
2-\alpha-\beta&1-\beta+\gamma& 1-\beta+\delta\ea; 1\right).\ee
However, a careful computation, exactly analogous to that for $\Phi_n(t)$, yields the result
\be\label{generalbasis2} \left( \tau^*\, \tau \, - n(n+\alpha+\beta+\gamma+\delta)\right)\Psi_n(t)=
\frac{\Gamma(2-\alpha-\beta)\Gamma(1-\beta+\gamma)\Gamma(1-\beta+\delta)}{\Gamma(1-n-\alpha-\beta)\Gamma(n+\gamma+\delta)
\Gamma(\alpha+t)\Gamma(\alpha-t)},\ee
so $\Psi_n(t)$ doesn't satisfy the eigenvalue equation. Now, comparing (\ref{generalbasis1}),(\ref{generalbasis2}),
we see that  functions 
\be\label{generalbasis3}{\tilde \Phi}_n^{(\alpha,\beta\gamma,\delta)}(t)=\Phi_n(t)-
\frac{\Gamma(\alpha+\beta)\Gamma(\alpha+\gamma)\Gamma(\alpha+\delta)\Gamma(1-n-\alpha-\beta)\Gamma(n+\gamma+\delta)}
{\Gamma(-n)\Gamma(n+\alpha+\beta+\gamma+\delta-1)\Gamma(2-\alpha-\beta)\Gamma(1-\beta+\gamma)\Gamma(1-\beta+\delta)}\Psi_n(t)\ee
do satisfy the eigenvalue equation for general $n=n'+c$ where $n'$ runs over the  integers and $c$ is a fixed noninteger. Furthermore it is straightforward 
to verify that  ${\tilde \Phi}_n^{(\alpha,\beta\gamma,\delta)}(t)$ satisfies all of the recurrence formulas induced by the intertwining operators 
$\tau,\mu , \tau^*,\mu^*$ for general $n$, such as 
(\ref{murec}), that are satisfied by $\Phi_n$ for nonnegative integer values. The functions ${\tilde \Phi}_n(t)$  have 
the duality property, hence since they satisfy the 2nd order difference eigenvalue equation  for general $n$ they must also 
satisfy the 3-term 
recurrence formula. Thus for fixed $c$ these basis functions define an infinite-dimensional irreducible representation of the quadratic algebra of 
$S9$ in which the eigenvalues of $L_1$ are indexed by an arbitrary integer $n'$. In the original quantum problem this representation 
is easy to construct: all of the recurrence relations that we have derived using hypergeometric functions 
remain valid for nonpolynomial hypergeometric functions. These nonpolynomial bases are no longer normalizable, but the recurrence relations 
remain valid. Not  clear was how such representations could be realized in terms of difference operators. The solution is that 
the analytic continuation of the  basis functions
is (\ref{generalbasis3}),  the Wilson functions which lead to associated Wilson polynomials \cite{Ismail, Masson}, 
and to  the Wilson 
transform \cite{Groenevelt}, which  corresponds 
to infinite dimensional irreducible representations of the quadratic algebra of $S9$ in which the spectrum of $L_1$ is (partly) continuous.
In the original quantum mechanical system the analytic continuation is evident, but the integral over the sphere, i.e. in $x,y$,  giving the 
interbasis expansion coefficients is deformed into a Pochhammer contour integral on a Riemann surface over the  $x$-plane with branch points at $\pm 1$ and a 
similar surface over the $y$-plane. All of the intertwining operator recurrences can be verified by integration by parts.

\section{Discussion and Conclusions}

\begin{itemize}
\item We showed explicitly how Racah and Wilson polynomials, and the Wilson functions, arise as expansion coefficients for the generic superintegrable system on the complex 2-sphere, relating two
 different sets of 
spherical coordinate bases. We employed only techniques from quantun theory and facts about the differential
recurrence relations of ordinary ${}_2F_1$ hypergeometric series. 
We did not use results from the established theory of Wilson polynomials and functions, although we did identify instances 
where our results correspond to known facts about the functions. 
\item We showed how the principal properties of these functions: the measure, 3-term recurrence relation, the orthogonality measure, 2nd order difference equation,
 duality, permutation symmetry, and intertwining operators --  follow from the symmetry of this quantum system. 
\item The parameter changing difference relations for the polynomials follow from intertwining operators for the quantum system. 
All of the properties of this
 system are induced by the fundamental differential
recurrence relations of the Gaussian hypergeometric functions. 

\item  There is active interest in the relation between multivariable Wilson polynomials and  the quantum superintegrable system on the $n$-sphere with
  generic  potential, e.g. \cite{KMP2011}, and these results should aid in the generalization.
\item By contracting function space realizations of irreducible representations of the $S9$  quadratic  algebra  to  the other
 superintegrable systems one obtains the 
full Askey scheme of orthogonal hypergeometric polynomials, uniquely induced
by  Lie algebra contractions of $so(3,C)$ and $e(2,C)$, \cite{KMP2014,KM2014}. This work should be extended to multivariable orthogonal polynomials.
\end{itemize}

\ack W.\, M.\, benefited greatly from conversations with Charles Dunkl, Mourad Ismail,  Eric Koelink and Tom Koornwinder. 
We thank the refferee for several important suggestions.This work was partially supported by a grant from the Simons Foundation (\# 208754 to Willard Miller, Jr.).

\end{document}